\documentclass[preprint,review,12pt]{elsarticle}

\usepackage{amsmath}
\usepackage{amssymb}
\usepackage{amsfonts}
\usepackage{graphicx}
\usepackage{dcolumn}
\usepackage{bm}
\usepackage{hyperref}
\hypersetup{colorlinks=false}
\biboptions{numbers,square}

\journal{Annals of Physics}

\begin{document}

\begin{frontmatter}

\title{Self-force on an electric dipole in the spacetime of a cosmic string}

\author{C. R. Muniz}

\address{Grupo de F\'isica Te\'orica (GFT), Universidade Estadual do Cear\'a, UECE-FECLI, Iguatu, Cear\'a, Brazil.}

\ead{celiomuniz@yahoo.com}

\author{V. B. Bezerra}

\address{Departamento de F\'{i}sica, Universidade Federal da Para\'{i}ba, Caixa Postal 5008, CEP 58051-970, Jo\~{a}o Pessoa, PB, Brazil}

\ead{valdir@ufpb.br}

\begin{abstract}
We calculate the electrostatic self-force on an electric dipole in the spacetime generated by a static, thin, infinite and straight cosmic string. The electric dipole is held fixed in different configurations, namely, parallel, perpendicular to the cosmic string and oriented along the azimuthal direction around this topological defect, which is stretched along the $z$ axis. We show that the self-force is equivalent to an interaction of the electric dipole with an effective dipole moment which depends on the linear mass density of the cosmic string and on the configuration.  The plots of the self-forces as functions of the parameter which determines the angular deficit of the cosmic string are shown for those different configurations.
\end{abstract}


\end{frontmatter}

\section{Introduction}
\label{1}
The phenomenon of self-interaction force experienced by an electric charge placed in an arbitrary gravitational field was analyzed for the first time by DeWitt and Brehme \cite{DeWitt}. In this scenario, the self-force is induced by the interaction of the electromagnetic field generated by the electric charge and the curvature of the spacetime \cite{DeWitt2, Hobbs, Smith, Vilenkin, Piazzese, Boisseau, Bakhmatov, Boisseau2}. This interaction modifies the electrostatic potential in such a way that the electric charge experiences a finite self-force. This is an example of how the self-force on a charged particle held fixed in a gravitational field depends on the local aspects (geometry) of the spacetime.

In fact, the self-interaction force depends not only on the geometry of the spacetime but also on the global features of the spacetime. This can be understood as a manifestation of non-local (topological) influence of gravity on the electromagnetic field, and means that non-local aspects of the gravitational fields are of fundamental importance in describing this physical system. In other words, the intrinsic geometry of the space is not sufficient to describe completely the interaction of the electromagnetic field produced by an electric charge and the gravitational field.

A scenario in which we can investigate the topological features of the interaction between an electromagnetic and a  gravitational field is the spacetime produced by a thin, infinite and straight cosmic string \cite{Vilenkin2,Hiscock,Linnet,Gott}, which looks like the direct product of the two-dimensional Minkowski space and a cone, and thus it is locally flat but not globally. The metric outside a cosmic string positioned along of the $z$ axis can be written, in cylindrical coordinates, as
\begin{equation}\label{01}
ds^2=dt^2-d\rho^2-\frac{\rho^2}{\nu^2}d\phi^2-dz^2,
\end{equation}
where the domain of the azimuthal angle is $0\leq\phi\leq2\pi$. The parameter $\nu$ is given by $\nu=(1-4G\mu)^{-1}$, where $\mu$ is the mass per unit length of the string. Note that we can redefine the azimuthal angle in such a way that $\theta=\frac{1}{\nu}\phi$, with $0\leq\theta<2\pi/\nu$. Thus, the metric expressed in eq. (\ref{01}) is actually a locally flat one, which means that the spacetime of an infinitely thin cosmic string may be regarded as the Minkowski spacetime with a deficit angle $2\pi(1-\nu^{-1})$.

This lack of global flatness of the cosmic string spacetime originates an electrostatic self-force on a charged particle, investigated in \cite{Linnet2}-\cite{valdir}, on a linear charge distribution \cite{Carvalho}, on a mean charge density \cite{Bezerra}, on an electric dipole in presence of a point mass and in the context of planar gravity \cite{grats}, or on an electric dipole held fixed parallel to the cosmic string \cite{Bezerra2,Barbosa}. The self-forces on a particle \cite{burko} and on a distribution of charges \cite{harte}, placed in other locally Minkowski space-times, also were analysed.

The purpose of this work is to investigate how the self-force phenomenon induced by the global features of the gravitational field of a cosmic string manifests itself on an electric dipole held fixed in different configurations, as for example, parallel, perpendicular to the cosmic string and oriented along the azimuthal direction (orthogonal), and in what way these forces depend on the parameter which defines the deficit angle of the spacetime under consideration.

A similar problem has been already investigated using the retarded Green's function \cite{pavel} associated with the electromagnetic field of a electric dipole and of other static distributions of charge, in the spacetime of a straight, thin and infinite cosmic string. It was shown that these sources are influenced by the cosmic string through a self-interaction with their fields. In our paper, the self-interaction on an electric dipole placed in the cosmic string spacetime was derived by a different method which was the same one used in \cite{Linnet2}.

This paper is organized as follows: In Sec. II we revise the renormalized Green's function method, necessary for the calculation of self-energy on an electric charge. In Sections III, IV and V we obtain the self-energy and use this result to calculate the self-force on a point electric dipole when it is parallel, perpendicular and orthogonal to the cosmic string, respectively. We also obtain for each configuration the graphs which describes how the self-force depends on the parameter which defines the deficit angle. We discuss our results in Sec.VI.

\section{The Regularized Green's Method Revisited}

 The volumetric charge distribution of an electric dipole with point charges $q$ and $-q$, separated by a distance $\textbf{a}$, is given by
\begin{equation}\label{02}
\rho(\textbf{x})=\rho_{q}(\textbf{x})+\rho_{-q}(\textbf{x})=q[\delta^{3}(\textbf{x}-\textbf{x}_{1})-\delta^{3}(\textbf{x}-\textbf{x}_{2})],
\end{equation}
where $\textbf{x}_{1}$ and $\textbf{x}_{2}$ are the positions of the charges. In terms of the electric dipole moment $\textbf{P}=q\textbf{a}$, we have
\begin{equation}\label{03}
\rho(\textbf{x})=-\textbf{P}\cdot\nabla_{x}\delta^{3}(\textbf{x}-\textbf{x}').
\end{equation}
The subscript in nabla operator designates derivative with respect to $\bold{x}$ coordinates, and $\bold{x}'$ are the coordinates which localize the position of the electric dipole. Thus, using the expression for the self-energy of a continuum system of charges,
\begin{equation}\label{04}
U_{self}=\frac{1}{2}\int\int d^{3}\textbf{x}'d^{3}\textbf{x}''\rho(\textbf{x}')\rho(\textbf{x}'')G(\textbf{x}',\textbf{x}''),
\end{equation}
and taking into account the appropriate renormalized Green's function for point charges, the self-energy of an electric dipole at the point $\textbf{x}$ is given by
\begin{eqnarray}\label{05}
U_{self}(\textbf{x})=\frac{1}{2}\int d^{3}\textbf{x}'\textbf{P}\cdot\nabla_{x'}\delta^{3}(\textbf{x}-\textbf{x}')\int d^{3}\textbf{x}''\textbf{P}\cdot\nabla_{x''}\delta^{3}(\textbf{x}-\textbf{x}'')G^{ren}(\textbf{x}',\textbf{x}''),
\end{eqnarray}
or, using the properties of the delta function, we get
\begin{equation}\label{06}
U_{self}(\textbf{x})=\frac{1}{2}\textbf{P}\cdot\nabla_{x'}\big\{[\textbf{P}\cdot\nabla_{x''} G^{ren}(\textbf{x}',\textbf{x}'')]_{x''=x}\big\}_{x'=x}.
\end{equation}
Using eq. (\ref{06}) we calculate the self-energy of an electric dipole held fixed in the conical geometry associated with the spacetime of a cosmic string. In order to obtain this result, we have to calculate the renormalized Green's function in the background spacetime of a cosmic string. The non-renormalized one in this spacetime is a solution of the three-dimensional Poisson's equation in cylindrical coordinates
\begin{equation}\label{07}
\nabla^2G(\textbf{x}',\textbf{x}'')=\frac{1}{|\rho'-\rho''|}\delta(\rho'-\rho'')\delta(z'-z'')\delta(\theta'-\theta''),
\end{equation}
with the domain of $\theta$ being $[0,2\pi/\nu]$. The renormalized Green's function is obtained by subtracting from solution of the equation (\ref{07}) the Green's function in Minkowski spacetime, which is also solution of this equation but with the angular coordinate ranging in the interval $[0,2\pi]$. The result is \cite{Linnet2,Anderson}

\begin{eqnarray}\label{08}
G^{ren}(\textbf{x}',\textbf{x}'')&=&\frac{1}{\pi\sqrt{2\rho'\rho''}}\bigg\{\int_{\eta}^{\infty}\frac{\nu\sinh(\nu u)[\cosh{u}-\cosh{\eta}]^{-1/2}}{\cosh(\nu u)-\cosh\nu(\theta'-\theta'')}du+\nonumber\\
&-&\int_{\eta}^{\infty}\frac{\sinh{u}[\cosh{u}-\cosh{\eta}]^{-1/2}}{\cosh{u}-\cosh(\theta'-\theta'')}du\bigg\}.
\end{eqnarray}

Now, we substitute eq.(\ref{08}) into eq.(\ref{06}) to obtain the self-energy. From the expression for the self-energy, we can obtain the self-force through the following relation
\begin{equation}\label{09}
\textbf{F}=-\textbf{$\nabla$}U_{self},
\end{equation}
where $\cosh{\eta}=[\rho'^2+\rho''^2+(z'-z'')]^2(2\rho'\rho'')^{-1}$ and $\nu=(1-4\mu G)^{-1}$.

\section{Self-Force on an electric dipole perpendicular to the cosmic string}

Firstly, let us consider that the electric dipole is oriented along the radial direction related to the cosmic string which we will take as held fixed along the $z$ axis. Thus, the self-energy given by ($\ref{06}$), turns into
\begin{equation}\label{10}
  U_{self}(\textbf{x})=\frac{1}{2}P^{2}\frac{\partial}{\partial\rho'}\bigg\{\left[\frac{\partial}{\partial\rho''}G^{ren}(\textbf{x},''\textbf{x}')\right]_{x''=x}\bigg\}_{x'=x},
\end{equation}
and then, we have
\begin{eqnarray}\label{11}
\left[\frac{\partial}{\partial\rho''}G^{reg}(\textbf{x}',\textbf{x}'')\right]_{x''=x}=-\frac{1}{\pi\sqrt{2\rho'}\rho^{3/2}}\int_{\eta}^{\infty}f_{\nu}(u,\eta)du+\frac{1}{\pi\sqrt{2\rho'\rho}}\frac{\partial\eta}{\partial\rho''}\frac{\partial}{\partial\eta}\int_{\eta}^{\infty}f_{\nu}(u,\eta)du,\nonumber\\
\end{eqnarray}
where $f_{\nu}(u,\eta)$ is the sum of integrands in the right hand side of eq. (\ref{08}). We also have that
\begin{equation}\label{12}
\left[\frac{\partial\eta}{\partial\rho''}\right]_{x''=x}=\frac{\rho^{2}-\rho'^{2}-z^{2}}{2\rho'\rho^{2}}\frac{1}{\sinh{\eta}}, \end{equation}
and
\begin{eqnarray}\label{13}
\frac{\partial}{\partial\eta}\int_{\eta}^{\infty}f_{\nu}(u,\eta)du=-\displaystyle\lim_{u\to \eta}f_{\nu}(u,\eta)+\int_{\eta}^{\infty}\frac{\partial f_{\nu}(u,\eta)}{\partial\eta}du.\nonumber\\
\end{eqnarray}
Applying the derivative with respect to $\rho'$ and taking the coincidence limit $\textbf{x}'=\textbf{x}$, the self-energy and the self-force on the electric dipole are given, respectively, by
\begin{subequations}
\begin{equation}\label{14}
U_{self}(\rho)=\frac{P^2}{4\pi\rho^{3}}[F(\nu)+G(\nu)],
\end{equation}
\begin{equation}
\label{14a}
F_{self}(\rho)=\frac{3P^2}{4\pi\rho^4}[F(\nu)+G(\nu)].
\end{equation}
\end{subequations}
where
\begin{eqnarray}\label{15}
F(\nu)=\bigg\{\int_{0}^{\infty}\frac{\nu\sinh(\nu u)[\cosh{u}-1]^{-1/2}}{\cosh(\nu u)-1}du-\int_{0}^{\infty}\frac{\sinh{u}[\cosh{u}-1]^{-1/2}}{\cosh{u}-1}du\bigg\},\nonumber\\
\end{eqnarray}
and
\begin{eqnarray}\label{16}
G(\nu)=\bigg\{\int_{0}^{\infty}\frac{\nu\sinh(\nu u)[\cosh{u}-1]^{-3/2}}{\cosh(\nu u)-1}du-\int_{0}^{\infty}\frac{\sinh{u}[\cosh{u}-1]^{-3/2}}{\cosh{u}-1}du\bigg\}.\nonumber\\
\end{eqnarray}

From eq. (14a) we conclude that the self-force can be seen as arising from a interaction of the real electric dipole with another effective dipole moment $P _{eff}=\frac{P}{16\pi}[F(\nu)+G(\nu)]$ collinear to the former and placed at the cosmic string position.

We depict in Fig.1 the self-force on the electric dipole in the perpendicular configuration, as a function of the parameter $\nu$. We can also verify from the graph that the (radial) self-force on the dipole is repulsive.

\begin{figure}[!ht]
\centering
\includegraphics[scale=0.9]{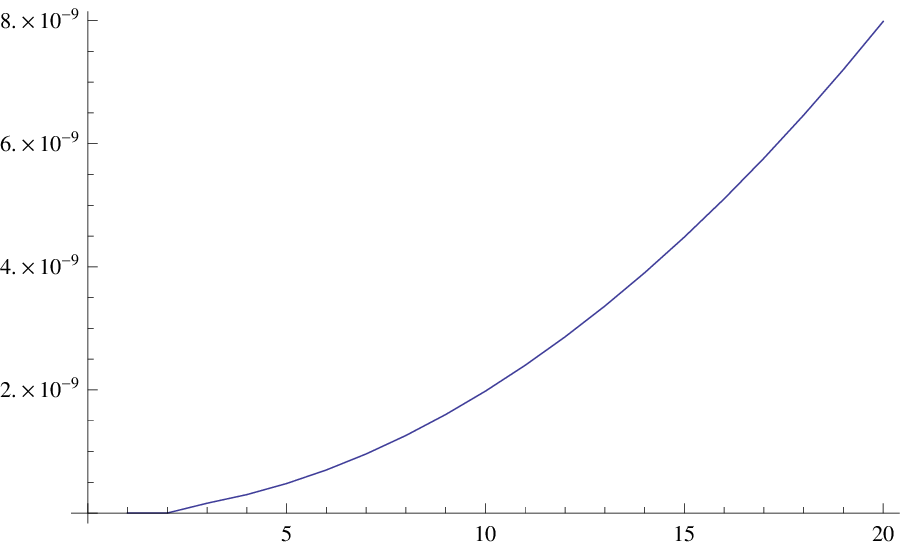}
\caption{Self-force on an electric dipole perpendicular to the cosmic string, as a function of angular parameter $\nu$, in dynes, for a radial distance $\rho=1cm$.}
\end{figure}

If we take into account that $G\mu\sim10^{-7}$, which is an upper limit imposed by recent cosmological observations \cite{keisler}, then $\nu\sim1.0+4.0\times10^{-7}$ and $U_{self}\sim-10^{-2}$ eV, for the radial distance of $1cm$.

\section{self-force on an electric dipole parallel to the cosmic string}

When the dipole is oriented along the direction parallel to the cosmic string, i.e., parallel to the $z$ axis, the self-energy given by eq. ($\ref{06}$) reduces to

\begin{equation}\label{16}
  U_{self}(\textbf{x})=\frac{1}{2}P^{2}\frac{\partial}{\partial z'}\bigg\{\left[\frac{\partial}{\partial z''}G^{reg}(\textbf{x},''\textbf{x}')\right]_{x''=x}\bigg\}_{x'=x}.
\end{equation}
Thus, we have
\begin{eqnarray}\label{17}
\left[\frac{\partial G^{reg}(\bf{x}',\bf{x}'')}{\partial z''}\right]_{x''=x}&=&\frac{-(z'-z)}{2\pi\sqrt{2\rho'}\rho^{5/2}}\bigg\{\int_{\eta}^{\infty}\frac{\nu\sinh(\nu u)[\cosh{u}-\cosh{\eta}]^{-3/2}}{\cosh(\nu u)-\cosh\nu(\theta'-\theta'')}du+\nonumber\\
&-&\int_{\eta}^{\infty}\frac{\sinh{u}[\cosh{u}-\cosh{\eta}]^{-3/2}}{\cosh{u}-\cosh(\theta'-\theta'')}du\bigg\}.
\end{eqnarray}
Taking the derivative with respect to $z'$ and considering the coincidence limit, we get, finally, the self-energy and the self-force, which are given by
\begin{subequations}
\begin{equation}\label{18}
U_{self}(\rho)=\frac{P^2}{4\pi\rho^3}G(\nu),
\end{equation}
\begin{equation}\label{18a}
F_{self}(\rho)=\frac{3P^2}{4\pi\rho^4}G(\nu).
\end{equation}
\end{subequations}
We depict in Fig.2 the self-force on the electric dipole in the parallel configuration, as a function of the parameter $\nu$. We can also verify from the graph that the (radial) self-force on the dipole is repulsive.
\begin{figure}[!ht]
\centering
\includegraphics[scale=0.9]{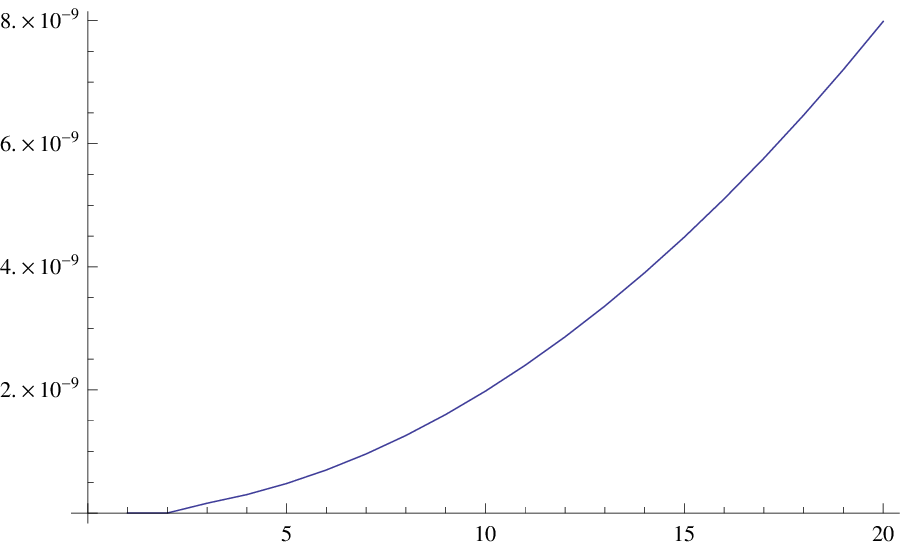}
\caption{Self-force on an electric dipole parallel to the cosmic string, as a function of angular parameter $\nu$, in dynes, for a radial distance $\rho=1cm$.}
\end{figure}

In the present case, the self-force can be also seen as resulting from an electrostatic interaction of the real electric dipole with an effective dipole moment $P _{eff}=\frac{P}{8\pi}G(\nu)$ parallel to the former and situated along the $z$ axis.

\section{self-force on an electric dipole orthogonal to the cosmic string}

Now, let us consider that the electric dipole is oriented along the azimuthal angle around the cosmic string. In this case, the self-energy is obtained from
\begin{equation}\label{19}
U_{self}(\textbf{x})=\frac{1}{2}\frac{P^2}{\rho^2}\frac{\partial}{\partial \theta'}\bigg\{\left[\frac{\partial}{\partial \theta''}G^{reg}(\textbf{x},''\textbf{x}')\right]_{x''=x}\bigg\}_{x'=x},
\end{equation}
whose result is more readly obtained than in the previously analysed cases, since that the derivative in eq. (\ref{19}) is not applied on the inferior limit of the integrals of Green's function expressed in equation (\ref{07}), because such limits depend on coordinates $\rho$ and $z$, only. Thus, the self-energy is simply
\begin{eqnarray}\label{20}
U_{self}=\frac{P^2}{2\pi\sqrt{2}\rho^3}H(\nu),
\end{eqnarray}
and the self-force is
\begin{equation}\label{21}
F_{self}=\frac{3P^2}{2\pi\sqrt{2}\rho^4}H(\nu).
\end{equation}
where
\begin{eqnarray}\label{22}
H(\nu)=\int_{0}^{\infty}\left\{\frac{\nu^3\sinh(\nu u)}{[\cosh(\nu u)-1]^2}
-\frac{\sinh{u}}{[\cosh{u}-1]^2}\right\}[\cosh{u}-1]^{-1/2}du.
\end{eqnarray}
In Fig.3, we depict the self-force for the configuration which we are considering in this section.
\begin{figure}[!ht]
\centering
\includegraphics[scale=0.9]{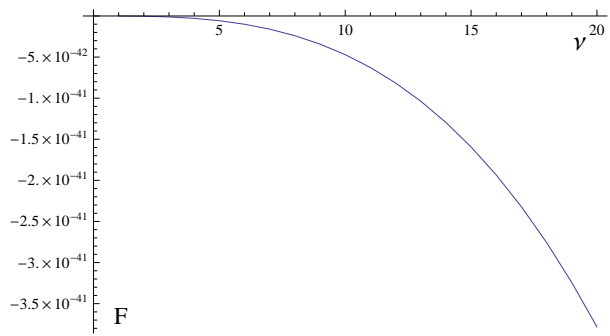}
\caption{Self-force on an electric dipole orthogonal to the cosmic string, as a function of angular parameter $\nu$, in dynes, for some radial distances.}
\end{figure}
 We note here great differences with respect to the other configurations. Firstly, the self-force on the orthogonal electric dipole is attractive. Secondly, the intensity of self-force is many orders of magnitude less than the observed in previous ones.

 The self-force also can be seen as the result of an interaction of the real electric dipole with another effective dipole  parallel to the former, placed at the position of the cosmic string, but with effective dipole moment given by $P _{eff}=\frac{P}{4\pi\sqrt{2}}H(\nu)$.

\section{Conclusions and Remarks}

We have calculated the self-interaction energy of a point electric dipole, placed at the spacetime of a straight, infinite, thin and static cosmic string, and used this result to find the self-force. These calculations were based on the renormalized Green's function formalism. We found that such self-forces are repulsive, for both perpendicular and parallel configurations, and attractive for the orthogonal one. In all cases, the self-force can be seen as a force exerted on the real electric dipole caused by the presence of an effective one held fixed at the cosmic string position, whose dipole moment depends on the characteristics of the cosmic string, codified in the parameter $\nu$ defined in terms of the linear mass density of the cosmic string. The resulting self-force, considering the three independent directions, is given by
\begin{equation}\label{23}
\textbf{F}_{self}=\frac{3P^2}{4\pi}\{[F(\nu)+G(\nu)]\textbf{e}_{\rho}+\sqrt{2}H(\nu)\textbf{e}_{\theta}+G(\nu)\textbf{e}_z\},
\end{equation}
and a general expression for the self-torque is
\begin{equation}\label{24}
\tau_i^{self}=-\epsilon_{ijk}P_j\frac{\partial U_{self}}{\partial P_k},
\end{equation}
where $\epsilon_{ijk}$ is the completely antisymmetric tensor in three spatial dimensions and $U_{self}$ is given by the sum of equations (14a), (19a) and (21).

We call attention to the fact that the total self-force on the electric dipole is not a mere superposition of self-forces on each particle that forms the dipole. The reason that the self-force of the dipole is not just a sum of self-forces of both charges forming it can be explained using the fact that, in the cosmic string spacetime, there is an additional interaction between both charges caused by the non-trivial topological properties of that spacetime.  

From the graphs that display the dependence of the self-forces on the parameter $\nu$, associated with the presence of the cosmic string, we found that the self-interaction forces on both parallel and perpendicular dipoles practically have the same values and that the self-force on the electric dipole aligned in the direction of the azimuthal angle is several orders of magnitude weaker as compared with the two other configurations. We consider in this analysis that each electric dipole is formed by elementary charges and that the distance between them is around $10^{-12}cm$, as well as that they are situated at a distance of $1cm$ from the cosmic string.

Beyond to investigate the interaction of cosmic strings with electric dipoles, which can be applied to study the interaction of complex molecules in interstellar and intergalactic media in order to detect the possible presence of cosmic strings, the results of this paper can be used, in principle, in condensed matter physics, through the analogy between cosmic strings and disclinations in solids, with the aim to understand the interaction of these defects with polar molecules.

\section*{Acknowleddgements}
C. R. Muniz would like to thank to Universidade Federal da  Para\'{i}ba for the kind welcome and to CNPq for a grant. V. B. Bezerra would like to thank CNPq for partial financial support.


\end{document}